\documentclass[11pt]{article}

\pdfoutput=1

\usepackage[T1]{fontenc}
\usepackage[latin9]{inputenc}
\usepackage[a4paper]{geometry}
\usepackage[active]{srcltx}
\usepackage{amsmath}
\usepackage{amssymb}
\usepackage{esint}
\usepackage{ulem}

\usepackage{xcolor}

\makeatletter

\usepackage{textcomp}

\pdfoutput=1 

\usepackage{jheppub}

\newcommand*\xbar[1]{%
  \hbox{%
    \vbox{%
      \hrule height 0.5pt 
      \kern0.3ex
      \hbox{%
        \kern-0.0em
        \ensuremath{#1}%
        \kern-0.0em
      }%
    }%
  }%
}

\usepackage{amsfonts}

\usepackage{bbm}

\newcommand{\be}{\begin{equation}}
\newcommand{\ee}{\end{equation}}
\newcommand{\bea}{\begin{eqnarray}}
\newcommand{\eea}{\end{eqnarray}}

\title{\boldmath Carroll supergravities}

\author[a,b]{Marc Henneaux}

\affiliation[a]{Universit\'e Libre de Bruxelles and International Solvay Institutes, \\ ULB-Campus Plaine CP231, B-1050 Brussels, Belgium}
\affiliation[b]{Coll\`ege de France,  Universit\'e PSL, 11 place Marcelin Berthelot, \\ 75005 Paris, France}

\emailAdd{marc.henneaux@ulb.be}

\preprint{}

\abstract{The electric and magnetic carrollian limits of $N=1$ supergravity in $D=4$ spacetime dimensions are explicitly derived.  The  approach is general and applies also to extended supergravity models. }

\makeatother

\begin{document}
\maketitle \flushbottom

\section{Introduction}

The Carrolian limits of Lorentz-invariant theories have attracted much interest in the past years in view of their potential applications (for recent reviews, see \cite{Bergshoeff:2022eog,Ruzziconi:2026bix}). In particular, two limits of Einstein theory have been considered in \cite{Henneaux:2021yzg}, the electric one and the magnetic one.  The purpose of this note is to construct the corresponding limits for supergravities.

The In\"on\"u-Wigner  contraction \cite{Inonu:1953sp} of the Poincar\'e algebra, obtained by rescaling the boost generators $K_i$ and the energy operator $H$ as $K_i \rightarrow \epsilon K_i$ and 
$H \rightarrow  \epsilon H$, can be extended to the Poincar\'e superalgebra by rescaling the spinor charges as $Q^A \rightarrow \sqrt{\epsilon} Q^A$ (see e.g. \cite{Koutrolikos:2023evq}).  This gives a supersymmetric extension of the Carroll algebra, in which the supercharges fulfill the (graded) commutator relations
\begin{eqnarray}
&& [M_{ij}, Q^A] = -  \frac{i}{2}  {(\Sigma_{ij})^A}_B Q^B \, , \label{Eq:TransSO(3)} \\
&& [K_i , Q^A] = 0   \label{Eq:InertBoost}\, ,\\
&& [Q^A, \bar{Q}_B] = - \frac{i}{2} {(\gamma_0)^A}_B  P^0  \, .
\end{eqnarray}
($[,]$ is here the graded commutator, i.e., the anticommutator for two fermionic objects such as the supercharges.)

These relations indicate in particular that in the Carroll limit,  the spinor charges do not transform under Carroll boosts. We will obtain the Carrollian supergravity theories by contraction of the corresponding Poincar\'e supergravities.  It therefore does not come as a surprise that in that approach, the Carroll spinors are automatically Carroll boost inert, for both the electric and magnetic limits.  

As indicated in \cite{Henneaux:2021yzg}, one quick way to derive the Carrollian limits is to perform the contraction within the Hamiltonian formulation.  This is because the contraction procedure treats differently the generators in space and time, so that the Hamiltonian is ``prepared" for taking the limits.

We  thus begin in Section {\bf \ref{Sec:Hamiltonian}} with a brief survey of the Hamiltonian form of  supergravity.  We  then derive in Section {\bf \ref{SEc:CarrollLimits}} the electric and magnetic Carrollian limits by appropriate contractions.   The electric limit is obtained in Subsection {\bf \ref{Subsec:Electric}}.  The magnetic limit is obtained in Subsection {\bf \ref{Subsec:Magnetic}}.  These are our main results.  We also focus on the properties of the corresponding constraint algebras, which are characteristic of the Carroll structure.  In  Section {\bf \ref{Sec:Covariant}} we rewrite the electric limit in covariant form and comment on the magnetic case.      Section {\bf \ref{Sec:Conclusions}} is devoted to concluding remarks.  Finally, a collection of appendices provide the necessary technical background:  Carroll spinors  are defined in Appendix {\bf {\ref{App:Spinors}}}; covariant derivatives of Lorentzian spinors and their decomposition in space and time are reviewed in Appendix {\bf \ref{Sec:LorentzianSpinors}};  Carrollian tetrads (``vierbein") and time gauge are discussed in Appendix {\bf \ref{Sec:CarrollTetrads}} ; and Appendix {\bf \ref{Sec:LieDerivatives}} provides information on Lie derivatives of spinor fieds.

\subsection*{The supergravity action}

Before we start, let us recall the form of the supergravity action.  This will give us the opportunity to fix our notations and conventions.

Greek indices run from 0 to 3 while Latin indices run from 1 to 3 (spatial indices).  Since we will have to distinguish between tetrad (`` vierbein")  components and components in the natural basis $\{\frac{\partial}{\partial x^\mu} \}$ (or their duals), we  shall systematically write the first ones in parenthesis. So, e.g., ${h^{(\alpha)}}_\mu$ is the $\mu$-component in the dual basis $\{ dx^\mu \}$ of the $1$-form $h^{(\alpha)}$.

The action for $N=1$ supergravity in $D=4$ spacetime dimensions reads \cite{VanNieuwenhuizen:1981ae,Freedman:2012zz}
\be
S = S_I + S_{II} + S_{III}
\ee
where $S_I[{h^{(\alpha)}}_\mu]$ is the Einstein action in tetrad form, 
\be
S_I[{h^{(\alpha)}}_\mu] = \int d  ^4 x \mathcal L_I \, , \quad \mathcal L_I = \frac{1}{2 \kappa^2} \, ^{4 \hspace{-.07cm}} R \sqrt{- ^{4 \hspace{-.04cm}} g} \, ,  \quad g_{\mu \nu} = \eta_{(\alpha) (\beta)} {h^{(\alpha)}}_\mu {h^{(\beta)}}_\nu \, , \quad \sqrt{- ^{4 \hspace{-.04cm}} g} = h 
\ee
($h = $det${h^{(\alpha)}}_\mu$), $S_{II}[\psi_\mu, {h^{(\lambda)}}_\mu]$ is the covariantized form of the Rarita-Schwinger action (with the tetrad-preserving, torsionless spin connection $\omega(h)$),
\be
S_{II}[\psi_\mu, {h^{(\alpha)}}_\mu] = \int d  ^4 x \mathcal L_{II} \, , \qquad \mathcal L_{II}  =  -  \sqrt{- ^{4 \hspace{-.04cm}} g} \, \bar{\psi}_{\mu}\gamma^{\mu \nu \sigma} D_\nu \psi_\sigma \, ,
\ee
and $S_{III}$ collects the quartic terms necessary for local supersymmetry invariance,
\begin{eqnarray}
&& S_{III}[\psi_\mu, {h^{(\alpha)}}_\mu] = \int d  ^4 x \mathcal L_{III} \, , \\
&& \mathcal L_{III} = 2 \kappa^2 \sqrt{- ^{4 \hspace{-.04cm}} g} \, L^{quartic} \, .
\end{eqnarray}
The term $L^{quartic}$ is purely algebraic and is explicitly given in \cite{VanNieuwenhuizen:1981ae,Freedman:2012zz}.  Its detailed expression will not be needed here but we shall rely on its well-established properties.

We use standard normalizations for the spin-3/2 field.  The constant $\kappa$ stands for
\be
\kappa^2 =\frac{8 \pi G}{c^4}
\ee
and is  kept since it will facilitate the derivation of the Carrollian limits.  

The action is invariant under supersymmetry,
\begin{eqnarray}
&&\delta {h^{(\alpha)}}_\mu = \frac12 \kappa^2 \bar \epsilon \gamma^{(\alpha)} \psi_\mu \, , \\
&& \delta \psi_\mu = ^{4 \hspace{-.2cm}}D_\mu \epsilon + \kappa^2 \epsilon \psi^2 \, ,
\end{eqnarray}
where $\psi^2$ stands for terms quadratic in the spinors, given explicitly in \cite{VanNieuwenhuizen:1981ae,Freedman:2012zz} but again not explicitly needed here.

\section{A brief derivation of the Hamiltonian formulation of  $N=1$, $D=4$ supergravity.}
\label{Sec:Hamiltonian}

The Hamiltonian form of supergravity has been derived soon after the theory was formulated \cite{Deser:1977ur,Pilati:1977ht}.  We first briefly go over the derivation, following the approach of \cite{Henneaux:1986cz} that shortcuts many steps of the standard Dirac procedure.

\subsection{First order formulation}

One can rewrite $S_I$ in first order form (in time derivatives) by introducing the auxiliary variable $P^{ij}$ following standard Hamiltonian lines, to get
\be
S_I[{h^{(\alpha)}}_m, P^{ij}; N, N^k] = \int d  ^4 x \left( P^{ij} \dot{g}_{ij} - N \mathcal E - N^k \mathcal P_k \right) 
\ee
with 
\be 
 \mathcal E = 2 \kappa^2 G_{ijmn} P^{ij}P^{mn} - \frac{1}{2 \kappa^2} R \sqrt{g} \, , \quad G_{ijmn} = \frac{1}{2 \sqrt{g}} (g_{im} g_{jn} + g_{in}g_{jm} - g_{ij} g_{mn})
\ee
and 
\be
\mathcal  P_k = - 2 {P_{\; \, k}^j}_{\vert j}
\ee
(all objects are now 3-dimensional and we performed the change of variables $({h^{(\alpha)}}_\mu) \rightarrow ({h^{(\alpha)}}_m, N, N^k)$ where $N$ and $N^k$ are respectively the lapse and the shift). If one extremizes the action $S_I$ with respect to the $P^{ij}$'s, one gets equations that can be solved for the $P^{ij}$'s themselves, namely,
\be
\dot{g}_{ij} = 4 \kappa^2 N G_{ijmn}P^{mn} + N_{i \vert j} + N_{j \vert i}  \, , \quad \Leftrightarrow  \quad P^{ij} = -\frac{1}{2 \kappa^2} G^{ijmn} K_{mn}
\ee
where $K_{ij}$ is the extrinsic curvature,
\be
K_{ij} = - \frac{1}{2N} (\dot{g}_{ij} - N_{i \vert j} - N_{j \vert i})
\ee
and 
\be G^{ijmn}= \sqrt{g} \left( \frac12 (g_{im} g_{jn} + g_{in} g_{jm}) - g_{ij} g_{mn} \right)
\ee
is  DeWitt's supermetric. 
These equations can therefore be used to eliminate them.  When this is done, one gets back the action $S_I$ in its original second order form, so that the $P^{ij}$'s are indeed auxiliary fields.

This property remains valid if one adds $S_{I}$ and $S_{II}$ to $S_I$, so that the total action $S_{I} + S_{II} + S_{III}$ is at this stage given by 
\be
S_I[{h^{(\alpha)}}_m, P^{ij}; N, N^k] + S_{II}[{h^{(\alpha)}}_m, \psi_{(\beta)} ; N, N^k] + S_{III}[ {h^{(\alpha)}}_m, \psi_{(\beta)} ; N, N^k]  \label{Eq:FirstOrderA}
\ee
with $P^{ij}$ appearing only in $S_I$.  Although this is not essential for taking the limit, we have made the change of variables $ \psi_{\mu} \rightarrow \psi_{(\beta)}$ because it will turn out to simplify the form of the Hamiltonian symplectic structure.

\subsection{Decomposition in space and time}

Since the action $S_I + S_{II} + S_{III}$ is linear in time derivatives, the remaining steps in the derivation of the Hamiltonian are straightforward.  We must redefine the variables so that the kinetic term takes the standard ``$p \dot{q}$" canonical form.  To achieve this task we first exhibit explicitly the time derivatives, by decomposing all covariant expressions into space and time.  We use from now on the time gauge where ${h^{(0)}}_m = 0$ \cite{Dirac:1962}.

Using the formulas given in Appendix {\bf \ref{Sec:LorentzianSpinors}}, one finds  that the supergravity Lagrangian $\mathcal L_{II}$  reads, in the time gauge,
\begin{eqnarray}
\mathcal L_{II} = && \sqrt{g} \bar{\psi}_{(a)} \gamma^{(0)} \gamma^{(a)(b)} \Delta_0 \psi_{(b)} \nonumber \\
&& -  N \sqrt{g} \bar{\psi}_{(a)} \gamma^{(0)} \gamma^{(a)(b)} \psi^{(c)} K_{(b)(c)} \nonumber \\
&& + N \sqrt{g}\left(\bar{\psi}_{(0)} \gamma_{(a)} \psi_{(b)} G^{(a)(b)(c)(d)} K_{(c)(d)} -2 \bar{\psi}_{(0)} \gamma^{(0)}\gamma^{(a)(b)} D_{(a)}\psi_{(b)} \right) \nonumber \\
&& -  N \sqrt{g} \left(\bar{\psi}_{(a)}  \gamma^{(a)(b)(c)} D_{(b)}\psi_{(c)} + D_{(a)} (\bar{\psi}_{(b)} \gamma^{(b)} \psi^{(a)})\right) \, .
\end{eqnarray}
The definition of the derivative $\Delta_0 \psi_{(a)}$ is recalled in Appendix  {\bf \ref{Sec:LorentzianSpinors}}, formula (\ref{Eq:DefDelta}).  This "extrinsic" derivative enjoys nice transformation properties and plays for the gravitino field $\psi_{(a)}$ the same role as the extrinsic curvature $K_{ij}$ plays for the spatial metric.

One can also decompose the quartic term into space and time as follows,
\be
\mathcal L_{III} = 2 \kappa^2 \left(\psi_{(0)}^A\psi_{(0)}^B (M_2)_{AB} + \bar{\psi}_{(0)} M_3 + M_4 \right) N \sqrt{g} \, ,
\ee
where $M_2$, $M_3$ and $M_4$ are respectively quadratic, cubic and quartic in the spatial  $\psi_{(m)}$'s ($A$, $B$ are spinor indices).

\subsection{Redefining the variables}

We now have everything at hand to complete the derivation of the Hamiltonian form of the action.

The spinor  Lagrangian $\mathcal L_{II}$ contains the extrinsic curvature and so modifies the momenta conjugate to the metric.  Explicitly, the terms containing (linearly) the time derivatives of the spatial metric $g_{ij}$, of the triads and of the gravitino field  (from both $\mathcal L_{I}$ and $\mathcal L_{II}$) are
\begin{eqnarray}
&& \left(P^{ij} + \frac12  \sqrt{g}\bar{\psi}_{(a)} \gamma^{(0)} \gamma^{(a)(b)} \psi^{(c)} {h_{(b)}}^{i} {h_{(c)}}^{j} - \frac12 \bar{\psi}_{(0)} \gamma_{(a)} \psi_{(b)}  {h^{(a)}}_{k} {h^{(b)}}_{l} G^{klij} 
\right) \dot{g}_{ij} \nonumber \\
&& \qquad + \sqrt{g} \bar{\psi}_{(c)} \gamma^{(0)} \gamma^{(c)(d)} \left(\dot{\psi}_{(d)}+ \frac12 \Xi_{(a)(b)} \Sigma^{(a)(b)} \psi_{(d)} - \psi_{(a)} {\Xi^{(a)}}_{(d)} \right) \, 
\end{eqnarray}
which contains antisymmetrized time derivatives of the triads through the quantity $ \Xi_{(a)(b)}$  defined in (\ref{Eq:DefXi}).

We thus make the change of variables  $P^{ij} \rightarrow \pi^{ij}$ with 
\be 
\pi^{ij} = P^{ij} + \frac12  \sqrt{g}\bar{\psi}_{(a)} \gamma^{(0)} \gamma^{(a)(b)} \psi^{(c)} {h_{(b)}}^{i} {h_{(c)}}^{j} - \frac12 \bar{\psi}_{(0)} \gamma_{(a)} \psi_{(b)}  {h^{(a)}}_{k} {h^{(b)}}_{l} G^{klij} \label{Eq:PiP}
\ee
as well as $\psi_{(0)} \rightarrow \chi$ with 
\be
\chi = N \psi_{(0)} \, .
\ee
The full action becomes
\begin{eqnarray} 
&& S[{h^{(a)}}_m,  \pi^{ij}, \psi_{(b)}; N, N^k, \chi] = I^{Kin} -I^{H} \\
&& \qquad I^{Kin} = \int d^4 x \left( \pi^{ij} \dot{g}_{ij} +\sqrt{g} \bar{\psi}_{(a)} \gamma^{(0)} \gamma^{(a)(b)} \dot{\psi}_{(b)} + \frac12 \Xi_{(a)(b)} J^{(a)(b)}_{\frac32} \right)\nonumber \\
&& \qquad I^H = \int d^4 x \left( N \mathcal H + N^k \mathcal H_k + \bar{\chi} \mathcal S \right)
\end{eqnarray}
where 
\be
 J^{(a)(b)}_{\frac32} =\sqrt{g} \bar{\psi}_{(c)} \gamma^{(0)} \gamma^{(c)(d)} \left(  \Sigma^{(a)(b)} \psi_{(d)} + \psi^{(b)}  \delta^{(a)}_{(d)} - \psi^{(a)}  \delta^{(b)}_{(d)} \right)
\ee
and where $I^{H}$ collects all the terms without time derivatives, to which we now turn.
 
As indicated above, the variables $P^{ij}$ should be viewed as the functions of the Hamiltonian variables defined by inverting (\ref{Eq:PiP}).   If one makes the substitution in $\mathcal E$, one gets terms quadratic in $\psi_{0}$, which exactly cancel the $M_2$ contribution, terms linear in $\psi_{(0)}$, which modify the generator $\mathcal S$ and terms independent of $\psi_{(0)}$, which redefine $M_4$.  The disappearance of the terms quadratic in $\psi_{(0)}$ is a consequence of the supersymmetry invariance of the action and has been verified in \cite{Deser:1977ur,Pilati:1977ht}.

Explicitly,  one finds for 
$ \mathcal H $, $\mathcal H_k$ and $\mathcal S$, 
\begin{eqnarray}
&& \mathcal H = 2 \kappa^2 G_{ijmn}\pi^{ij} \pi^{mn} - \frac{1}{2 \kappa^2} R \sqrt{g} \nonumber \\
&& \qquad \qquad -2 \kappa^2 \sqrt{g} G_{ijmn}\pi^{ij} \bar{\psi}_{(a)} \gamma^{(0)} \gamma^{(a)(b)} \psi^{(c)} {h_{(b)}}^{m} {h_{(c)}}^{n}+ \mathcal H_{\frac32}  \\
&& \mathcal H_{\frac32} =   \sqrt{g}\left(\bar{\psi}_{(a)}  \gamma^{(a)(b)(c)} D_{(b)}\psi_{(c)} + D_{(b)} (\bar{\psi}_{(a)} \gamma^{(a)}  \psi^{(b)}) \right) - 2 \kappa^2  \sqrt{g}M'_4 \\
&&M'_4 = M_4 -  \frac 14 G_{(a)(b)(c)(d)}    \bar{\psi}_{(e)} \gamma^{(0)} \gamma^{(e)(a)} \psi^{(b)}      \bar{\psi}_{(f)} \gamma^{(0)} \gamma^{(f)(c)} \psi^{(d)}  \\
&& \mathcal S  = 2 \kappa^2   \gamma_{(a)} \psi_ {(b)}  {h^{(a)}}_{m} {h^{(b)}}_{n} \pi^{mn}+ 2 \sqrt{g} \gamma^{(0)}\gamma^{(a)(b)} D_{(a)}\psi_{b} - 2 \kappa^2 \sqrt{g} M'_3 \\
&& M'_3 = M_3 + \frac 12 \sqrt{g} \gamma_{((a)}\psi_{(b))}\bar  \psi_{(c)} \gamma^{(c)((a)} \psi^{(b))} \, , \\
&& \int d^3 x N^k \mathcal H_k = \int d^3 x \left( \pi^{ij} \mathcal L_{N^k} g_{ij} + \sqrt{g} \bar{\psi}_{(a)} \gamma^{(0)} \gamma^{(a)(b)} \mathcal L_{N^k} \psi_{(b)}\right)
\end{eqnarray}
where the second sets of parentheses in $M'_3$ denotes symmetrization with weight one with respect to the indices $a$, $b$.

\subsection{The Hamiltonian form of the action}

In order to reach the complete Hamiltonian form, we need to treat the coefficients of the time derivatives ${h^{(a)}}_i$ in the kinetic term, i.e., 
\be
{\pi_{(a)}}^i = 2 \pi^{ij} h_{(a)j} - \frac12  J^{\frac32}_{(a)(b)} h^{(b)i}  \label{Eq:ConjugatePi}
\ee
 as independent variables.  Since $\pi^{ij}$ is symmetric, it follows from this relation that 
\be
J^{(a)(b)} \equiv J_{2}^{(a)(b)} + J_{\frac32}^{(a)(b)}  = 0 \, , \qquad J_{2}^{(a)(b)}\equiv \pi^{(a)i} {h^{(b)}}_i-  \pi^{(b)i} {h^{(a)}}_i
\ee
Conversely, if ${\pi_{(a)}}^i$ fullfills the constraint $J^{(a)(b)} = 0$, then one can compute $\pi^{ij}$ from (\ref{Eq:ConjugatePi}) to get
\be
\pi^{ij} = \frac{1}{4} ({\pi_{(a)}}^i h^{(a)j} + {\pi_{(a)}}^j h^{(a)i})
\ee
  It follows that one can treat the variables ${\pi_{(a)}}^i$ as independent variables provided one takes into account the constraints $J^{(a)(b)} = 0$ with the help of  Lagrange multipliers, which we call $\lambda_{(a) (b)}$.

The variational principle is therefore equivalent to
\begin{eqnarray} 
&& S[{h^{(a)}}_m,  {\pi_{(b)}}^r, \psi_{(c)}; N, N^k, \chi, \lambda_{(a)(b)}] =S^{Kin} -S^{H} \\
&& \qquad S^{Kin} = \int d^4 x \left( {\pi_{(a)}}^r {\dot{h}^{(a)}}_r +\sqrt{g} \bar{\psi}_{(a)} \gamma^{(0)} \gamma^{(a)(b)} \dot{\psi}_{(b)}  \right)\nonumber \\
&& \qquad S^H = \int d^4 x \left( N \mathcal H + N^k \mathcal H_k + \bar{\chi} \mathcal S + \frac12 \lambda_{(a)(b)} J^{(a)(b)}\right)
\end{eqnarray}
This is the action in canonical form, with canonical variables ${h^{(a)}}_m$,  ${\pi_{(b)}}^r$, $\psi_{(c)}$ and Lagrange multipliers $N$, $N^k$, $\chi$, $\lambda_{(a)(b)}$.

From the kinetic term, one reads that ${h^{(a)}}_m$  and ${\pi_{(b)}}^r$ are canonically conjugate,
\be
\{ {h^{(a)}}_m(x),{\pi_{(b)}}^r(y) \} = \delta^{(a)}_{(b)} \delta^r_m \delta(x,y)
\ee 
while the spinors are self-conjugate in the sense that
\be
\{ \psi_{(a)}^A(x), \psi_{(b)}^B(y)\} = - \frac{i}{4} g^{-\frac12} (\gamma_{(b)} \gamma_{(a)})^{AB} \delta(x,y) \, .
\ee
($\{, \}$ is the graded Poisson bracket and $A$, $B$ are spinor indices.)

Since the kinetic term of the gravitino involves the metric, the bracket of $\psi_{(a)} (x)$ with ${\pi_{(b)}}^r(y)$ does not vanish.  Working with the local fields $\psi_{(a)}$ rather than  $\psi_m$ is a first step towards achieving the canonical form of the brackets, which must be completed by considering the weighted spinor field $ \phi_{(a)} = g^{\frac14}\psi_{(a)}$ \cite{Deser:1977ur,Tabensky:1977ic}, for which one has  $\{ \phi_{(a)}^A(x), {\pi_{b}}^i(y)\} = 0$ (and $\{ \phi_{(a)}^A(x), \phi_{(b)}^B(y)\} = - \frac{i}{4}  (\gamma_{(b)} \gamma_{(a)})^{AB} \delta(x,y)$).

The constraints $\mathcal H \approx 0$, $ \mathcal H_k \approx 0$, $ \mathcal S \approx 0$ and $ J^{(a)(b)} \approx 0$ are all first class. Each constraint term generates a gauge transformation: $\mathcal H$ generates normal displacements to the equal time hypersurfaces,  $ \mathcal H_k$ generates tangential displacements, $ \mathcal S  $ generates local supersymmetry while $ J^{(a)(b)}$ generates local rotations of the triads.  That the constraints are first class is an automatic consequence of the gauge invariances of the supergravity action.

\section{Carrollian limits}
\label{SEc:CarrollLimits}

One convenient property of the time gauge is that its residual local subgroup is $so(3)$, for both the Lorentzian and Carrollian cases, so that the translation rules are direct in that gauge.

\subsection{Electric limit}
\label{Subsec:Electric}

The electric limit is obtained by rescaling $N$ and $\chi$ as 
\be 
N^E = 2 \kappa^2 N \, , \chi^E = 2 \kappa^2 \chi \, , 
\ee
and taking the limit $\kappa^2 \rightarrow \infty$, keeping $N^E$ and the other independendent variables fixed.

The action becomes 
\begin{eqnarray} 
&& S^E[{h^{(a)}}_m,  {\pi_{(b)}}^r, \psi_{(c)}; N^E, N^k, \chi^E, \lambda_{(a)(b)}] =S^{Kin} -S^{H, E} \\
&& \qquad S^{Kin} = \int d^4 x \left( {\pi_{(a)}}^r {\dot{h}^{(a)}}_r +\sqrt{g} \bar{\psi}_{(a)} \gamma^{(0)} \gamma^{(a)(b)} \dot{\psi}_{(b)}  \right)\nonumber \\
&& \qquad S^{H,E} = \int d^4 x \left( N^E \mathcal H^E + N^k \mathcal H_k + \bar{\chi}^E \mathcal S^E + \frac12 \lambda_{(a)(b)} J^{(a)(b)}\right)
\end{eqnarray}
where now
\begin{eqnarray}
&& \mathcal H^E =  G_{ijmn}\pi^{ij} \pi^{mn}
 - \sqrt{g} G_{ijmn}\pi^{ij} \bar{\psi}_{(a)} \gamma^{(0)} \gamma^{(a)(b)} \psi^{(c)} {h_{(b)}}^{m} {h_{(c)}}^{n}+ \mathcal H_{\frac32}  \\
&& \mathcal H_{\frac32} =  -  \sqrt{g}M'_4 \\
&& \mathcal S^E  =    \gamma_{(a)} \psi_ {(b)}  {h^{(a)}}_{m} {h^{(b)}}_{n} \pi^{mn} -  \sqrt{g} M'_3  \, ,
\end{eqnarray}
where $M'_4$ and $M_3$ are unaffected by the limit and takes the same form as in the Lorentzian case.

Except in the kinematical constraint $\mathcal H_k$, which generates spatial Lie derivatives, all spatial gradients have disappeared.  The quartic terms, which are algebraic, remain. One obtains $\mathcal H^E$ and $\mathcal S^E$ from the corresponding generators of the pseudo-riemannian case by simply dropping spatial derivatives. This makes these generators ``ultralocal".

We stress that the gauge invariance of the action and the first class property of the constraints is guaranteed to hold because it holds for all values of $2 \kappa^2$ and the limit $2 \kappa^2 \rightarrow \infty$ encounters no singularity.  So there is nothing to be checked in that respect.

\subsection{Magnetic limit}
\label{Subsec:Magnetic}

The magnetic limit is obtained by rescaling $N$  as 
\be 
N^M = (2 \kappa^2)^{-1} N \, , 
\ee
and taking the limit $\kappa^2 \rightarrow 0$, keeping $N^M$ and the other independendent variables fixed.  One may also view this limit as a zero-speed-of-light-limit $c \rightarrow 0$ if one rescales $G$ to zero faster than $c^4$.

The action greatly simplifies in that case to
\begin{eqnarray} 
&& S^M[{h^{(a)}}_m,  {\pi_{(b)}}^r, \psi_{(c)}; N^M, N^k, \chi, \lambda_{(a)(b)}] =S^{Kin} -S^{H, M} \\
&& \qquad S^{Kin} = \int d^4 x \left( {\pi_{(a)}}^r {\dot{h}^{(a)}}_r +\sqrt{g} \bar{\psi}_{(a)} \gamma^{(0)} \gamma^{(a)(b)} \dot{\psi}_{(b)}  \right)\nonumber \\
&& \qquad S^{H,M} = \int d^4 x \left( N^M \mathcal H^M + N^k \mathcal H_k + \bar{\chi} \mathcal S^M + \frac12 \lambda_{(a)(b)} J^{(a)(b)}\right)
\end{eqnarray}
where now
\begin{eqnarray}
&& \mathcal H^M =  -  R \sqrt{g} \\
&& \mathcal S^M  =  2 \sqrt{g} \gamma^{(0)}\gamma^{(a)(b)} D_{(a)}\psi_{(b)} \, .
\end{eqnarray}

Spatial gradients, both of the metric (in $\mathcal H$) and of the gravitino field (in $\mathcal S$)  remain in the magnetic limit, but quartic terms have gone.  Again gauge invariance is automatic so that it does not need to be checked.

As for the pure gravity theory in the magnetic limit, the equation of motion obtained by varying the conjugate momenta $\pi^{ij}$ cannot be solved for the momenta and force instead the condition that the extrinsic curvature should vanish.

\subsection{Constraint algebra}

The derivation of the Carrollian limits of supergravity constitutes our central result. Although it might be somewhat involved by other methods, it is direct in the Hamiltonian formulation, which provides a powerful approach to Carrollian limits - once one has indeed the Hamiltonian formulation at hand.

The algebra of the constraints becomes extremely simple in the Carrollian limit.  The key relations are
\be
\{\mathcal S_A^E (x), \mathcal S_B^E (y) \} = - i\delta_{AB} \mathcal H^E  \delta(x-y) \, \qquad \{\mathcal H^E (x), \mathcal H^E (y) \} = 0
\ee
and similarly
\be
\{\mathcal S_A^M(x), \mathcal S_B^M (y) \} = - i \delta_{AB}\mathcal H^M  \delta(x-y) \, \qquad \{\mathcal H^M (x), \mathcal H^M (y) \} = 0
\ee
(without $\mathcal H_k$ in the right-hand sides)

They express that the supersymmetry generators are the square roots of the Hamiltonian generator $\mathcal H$, in the same way as they are the square roots of $\mathcal H$ and $\mathcal H_k$ in standard supergravity \cite{Teitelboim:1977fs}.

It is interesting to note that while both the electric and magnetic limits lead to the same constraint algebra, they focus on different aspects of the full supergravity theory.  In the electric case where the limit is ultralocal, it is the square root aspect of the quadratic term in the momenta that is kept, much in the same way as the fermionic constraint for the Dirac particle is the square root of the mass-shell condition.   In the magnetic case, the properties that are retained are those underlying the supergravity-based proof of the positivity of the energy theorem for gravity \cite{Deser:1977hu,Witten:1981mf}.

In fact, an interesting feature of the rigid supersymmetry algebra that remains  true in the Carrollian limit is that it implies positivity of the energy,
\be
[Q^A, \bar{Q}_B] = - \frac{i}{2} {(\gamma_0)^A}_B  P^0   \quad \Leftrightarrow [Q^A, Q^\dagger_B] =  \frac{1}{2}  P^0 \quad \Rightarrow \; 0 \leq Tr (Q Q^\dagger + Q^\dagger Q) = 2 P^0
\ee

In the magnetic case, where the dynamical constraints $\mathcal S$ and $\mathcal H$ involve spatial gradients, the dynamical  Carroll generators are given on-shell  by non identically vanishing surface integrals \cite{Perez:2021abf}.  Similar arguments can then be given that the surface integral for the energy is positive.  In the electric case, the surface integrals for the supercharge and the energy are zero, so positivity becomes the uninteresting statement $ 0 \geq 0$.

\section{Electric limit in covariant form}
\label{Sec:Covariant}

\subsection{First order action}

Using the geometric concepts developed in \cite{Henneaux:1979vn}, it is easy to rewrite the electric supergravity action in manifestly covariant form.   

The fastest  way to get the covariant form of the action is to start from the $\kappa^2 \rightarrow \infty$ limit of the first order action (\ref{Eq:FirstOrderA}) decomposed in space and time but in which the redefinition $P^{ij} \rightarrow \pi^{ij}$ has not been performed yet. 
Together with the rescaling $N^E = 2 \kappa^2 N $, it is convenient to set
\be
N^E K^E_{ij} = N K_{ij} =  -\frac12 \Delta_0 g_{ij} \, , \qquad \Delta_0 g_{ij} \equiv\dot{g}_{ij} - N_{i \vert j} - N_{j \vert i}
\ee
Both $K^E_{ij}$ and $\Delta_0 g_{ij}$ are quantities that remain finite in the electric carrollian limit (while $K_{ij}$ diverges since $N \rightarrow 0$).

Expanding the action in inverse  powers of $2 \kappa^2$ (working with $\psi_{(0)} = \chi^E /N^E$ which remains finite), we obtain
\be
S_I[{h^{(a)}}_m, P^{ij}, \psi_{(b)}; N^E, N^k, \psi_{(0)}] = \int d  ^4 x \left(\mathcal L^E_{(0)} + \frac{1}{2 \kappa^2} \mathcal L^E_{(1)} + \frac{1}{(2 \kappa^2)^2} \mathcal L^E_{(2)} \right) 
\ee
with
\begin{eqnarray}
 \mathcal L^E_{(0)} = &&  P^{ij} \dot{g}_{ij} - N^E G_{ijmn} P^{ij}P^{mn} +2 N^k {P_{\; \, k}^j}_{\vert j} +  \sqrt{g} \bar{\psi}_{(a)} \gamma^{(0)} \gamma^{(a)(b)} \Delta_0 \psi_{(b)} \nonumber \\
&& +  N^E \sqrt{g} \left(- \bar{\psi}_{(a)} \gamma^{(0)} \gamma^{(a)(b)} \psi^{(c)} K^E_{(b)(c)} +   \bar{\psi}_{(0)} \gamma_{(c)} \psi_{(d)} G^{(c)(d)(a)(b)} K^E_{(a)(b)} + L^{quartic} \right)\, , \qquad
\end{eqnarray}
\begin{eqnarray}
\mathcal L^E_{(1)} =  && - 2N^E \sqrt{g}  \bar{\psi}_{(0)} \gamma^{(0)}\gamma^{(a)(b)} D_{(a)}\psi_{(b)} \nonumber \\
&& -2N^E \sqrt{g}\bar{\psi}_{(a)} \left( \gamma^{(a)(b)(c)} D_{(b)}\psi^{(c)} - D_{(a)} (\bar{\psi}_{(b)} \gamma^{(b)} \psi^{(a)})\right)
\end{eqnarray}
and 
\be
\mathcal L^E_{(2)} = N^E \sqrt{g} R \, .
\ee
Thus, one gets in the limit $2 \kappa^2 \rightarrow \infty$,
\be
S_C[{h^{(a)}}_m, P^{ij}, \psi_{(b)}; N^E, N^k, \psi_{(0)}] = \int d  ^4 x \mathcal L^E_{(0)} 
\ee

For all values of $\kappa^2$, the action is invariant under  supersymmetry transformations, which read after redefinition of the supersymmetry parameter as 
\be
\epsilon^E = \kappa^2 \epsilon
\ee
to eliminate the divergences pieces in the limit,
\begin{eqnarray}
&&\delta {h^{(a)}}_m = \frac12  \bar \epsilon^E \gamma^{(a)} \psi_m \, , \\
&&\delta N^E =  \frac{N^E}{2} \bar \epsilon^E \gamma^{(0)} \psi_{(0)} \, , \\
&&\delta N^k =\frac{N^E}{4 \kappa^2}  \left(\bar \epsilon^E \gamma^{(a)} \psi_{(0)} + \bar \epsilon^E \gamma_{(0)} \psi^{(a)}\right){h_{(a)}}^k \, , \\
&& \delta \psi_{(a)} =  \frac{1}{\kappa^2}D_{(a)} \epsilon^E +  2K^E_{kn} {h_{(a)}}^k {h_{(b)}}^n \Sigma^{(0)(b)} \epsilon^E + \epsilon^E \psi^2 \, , \\
&& \delta \psi_{(0)} = \frac{2}{N^E}\Delta_0 \epsilon^E - \frac{1}{\kappa^2}  \frac{\partial_n N^E}{N^E} \Sigma^{(0)(b)} \epsilon^E {h_{(b)}}^n+  \epsilon^E \psi^2 \, , 
\end{eqnarray}
(see Appendix {\bf \ref{Subsec:SusyTimeGauge}}).

We can safely take the limit $2 \kappa^2 \rightarrow \infty$, yielding  the supersymmetry transformations,
\begin{eqnarray}
&&\delta {h^{(a)}}_m = \frac12  \bar \epsilon^E \gamma^{(a)} \psi_m \, , \\
&&\delta N^E =  \frac{N^E}{2} \bar \epsilon^E \gamma^{(0)} \psi_{(0)} \, , \\
&&\delta N^k =0 \, , \\
&& \delta \psi_{(a)} =    2K^E_{kn} {h_{(a)}}^k {h_{(b)}}^n \Sigma^{(0)(b)} \epsilon^E + \epsilon^E \psi^2  , \\
&& \delta \psi_{(0)} = \frac{2}{N^E}\Delta_0 \epsilon^E +  \epsilon^E \psi^2 \, , 
\end{eqnarray}
under which the (electric) Carrollian action $S_C[{h^{(a)}}_m, P^{ij}, \psi_{(b)}; N^E, N^k, \psi_{(0)}] $ is invariant.

\subsection{Covariant form of the action}
The gravitational part of the action $P^{ij} \dot{g}_{ij} - N^E G_{ijmn} P^{ij}P^{mn} +2 N^k {P_{\; \, k}^j}_{\vert j}$ can be easily covariantized.  By eliminating the variables $P^{ij}$ using their own equations of motion, one gets the second order action for the electric version of Carroll gravity \cite{Henneaux:1979vn} in tetrad form,
\be
S^E_2 [{h^{(\lambda)}}_\mu] = \int d^4 x h  (K^E_{\alpha \beta} K^{E \, \alpha \beta} - (K^E)^2) 
\ee
where
\be
K^E_{\alpha \beta} = -\frac12 \mathcal L_n g_{\alpha \beta}
\ee
and $K^E= {K^{E \, \alpha}}_\alpha$, $h =  \det \left({h^{(\alpha)}}_\mu \right) $.  The Carrollian `` extrinsic curvature" $K^E_{\alpha \beta}$ is transverse and its spatial components coincide with $K^E_{ij}$ so that $K^E_{\alpha \beta} K^{E \, \alpha \beta} = K^E_{ij} K^{E \, ij}$ and ${K^{E \, \alpha}}_\alpha = {K^{E \, i}}_i$.  Note that the expression $K^E_{\alpha \beta} K^{E \, \alpha \beta} - (K^E)^2$ is invariant under local Carroll transformations of the tetrads, so that we do not need to assume the time gauge.

To understand the covariance of the spinor part of the action, we need to prescribe first the transformation properties of the gravitino field under all Carroll transformations.  Since the Carroll spinors are inert under Carroll boosts, we shall impose similar properties on the vector index of $\psi_{(\mu)}$.  That is, we assume that $\psi_{(m)}$ defines a transverse field $\Phi_{(\mu)}$ ($\Phi_{(\mu)} n^{(\mu)} = 0 \Rightarrow \phi_{(0)} = 0$) such that $\Phi_{(m)} = \psi_{(m)}$.  And we also assume  that $\psi^{(0)}$ defines a longitudinal field $\Psi^{(\mu)} = \Psi n^{(\mu)} \Rightarrow \Psi^{(m)} = 0$ such that $\Psi = \psi^{(0)}$.

We then observe that the Lie derivative of $\Phi_{(\alpha)}$ along $n^\mu$, defined in Appendix {\bf \ref{Sec:LieDerivatives}}, is also transverse and such that
\be 
N^E \mathcal L_n \Phi_{(a)} = \Delta_0 \psi_{(a)} \, .
\ee
We can therefore rewrite the spinor of action in covariant form as
\begin{eqnarray}
S^E_{\frac32} [{h^{(\alpha)}}_\mu, \Phi_{(\beta)}, \Psi] = && \int d^4 x h \Big(\bar{\Phi}_{(\alpha)} \Gamma \rho^{(\alpha)(\beta)} \mathcal L_n \Phi_{(\beta)} -   \bar{\Phi}_{(\alpha)} \Gamma \rho^{(\alpha)(\beta)} \Phi^{(\gamma)} K^E_{(\beta)(\gamma)} \nonumber \\
&& \qquad \qquad \qquad-\bar{\Psi} \rho_{(\gamma)} \Phi_{(\delta)} G^{(\gamma)(\delta)(\alpha)(\beta)} K^E_{(\alpha)(\beta)} + L^{quartic} \Big)
\end{eqnarray}
where we have explicitly used the  Carroll gamma-matrices $\Gamma$, $\rho_{(\mu)}$ of Appendix {\bf \ref{App:Spinors}}.  Because $G^{(\gamma)(\delta)(\alpha)(\beta)}$ is contracted with transverse objects, the expression involving it in the action is well-defined. The same holds true for all indices contractions, which make sense because the objects involved are transverse. Again, just as for $S^E_{2}$, there is no reference to the time gauge in this expression for $S^E_{\frac32}$, which is thus generally valid.  

Note that one might define a covariant derivative along the normal $\nabla_n \Phi_{(\beta)} \equiv  \mathcal L_n \Phi_{(\beta)} -   \Phi^{(\gamma)} K^E_{(\beta)(\gamma)}$  in order to regroup the first two terms in the action.

\subsection{Remarks on the magnetic case}

As in the pure gravity case, writing the magnetic limit in manifestly covariant form, is more intricate.  One reason is that the limit involves spatial covariant derivatives, which exist only because the extrinsic curvature $K_{\alpha \beta}$ vanishes in the magnetic limit.  However, the vanishing of $K_{\alpha \beta}$ is an on-shell statement and of course, one needs to write the action off-shell.  A second, related, difficulty is that the limit of the supersymmetry transformations blows up off-shell unless one uses the equation of motion $K_{\alpha \beta} = 0$ to guarantee their finiteness.  This means that in order to get a smooth limit, one needs first to redefine the supersymmetry transformations by on-shell trivial transformations.

We expect that all these difficulties are overcome in the approach based on the gauging of the super-Carroll algebra, just as in the magnetic Carroll limit of pure gravity \cite{Campoleoni:2022ebj} (see also \cite{Hartong:2015xda,Bergshoeff:2017btm,Figueroa-OFarrill:2022mcy}).  The momentum conjugate to the graviton, which forces $K_{\alpha \beta} = 0$ on-shell, would then appear as the arbitrary transverse tensor that characterizes (torsion-free) Carroll connections when they exist.  It is hoped to return to this question in the future.

\section{Conclusions and comments}
\label{Sec:Conclusions}

In this paper, we have derived the Carrollian limits of $D=4$, $N=1$ supergravity.  The procedure is  simple and direct in the Hamiltonian formalism, where it also manifestly preserves the gauge symmetries of the theory (in their Carrollian version). In particular, the electric limit is an extension to supergravity of the zero signature limit introduced in  \cite{Teitelboim:1980}.

The method can be applied to extended supergravity models, for which the Hamiltonian and the limits can be derived along the same lines (see for instance \cite{Henneaux:1986cz} for the Hamiltonian formulation of the $10$-dimensional models).

We have seen that the supersymmetry generators fullfill $\{ \mathcal S^A(x) , \mathcal S^B(y)\} = - i \delta^{AB}\mathcal H \delta(x-y)$, which is the local version of the (anti-)commutation relations $[Q^A, \bar{Q}_B] = - \frac{i}{2} {(\gamma_0)^A}_B  P^0$ of the global charges guaranteeing positivity of the energy.  Although written for the local generators, we expect that these relations should also be fulfilled by the $0$-th component of the supersymmetry current and the energy density  of a (Carroll) supersymmetric theory, paralleling the properties of the commutation relations of the energy density at different spacelike-related points exhibited for a Poincar\'e-invariant theory in \cite{Dirac:1962aa,Schwinger:1963xx}.
These commutation relations might be a tool to explore supersymmetric Carroll theories, and in particular ``Carroll swiftons", i.e., Carroll tachyons with positive energy \cite{Ecker:2024czx}.  The use of supersymmetry should also have an impact on the unusual features encountered in the quantization of Carroll field theories
\cite{deBoer:2023fnj}.

\section*{Acknowledgements}
 This work was partially supported by FNRS-Belgium (convention IISN 4.4503.15), as well as by research funds from the Solvay Family.

\appendix

\section{Lorentz and Carroll Spinors}
\label{App:Spinors}

Carroll spinors have been considered previously in \cite{Bagchi:2022eui,Bagchi:2022owq,Bergshoeff:2024ytq} so that we give here only a brief survey (along somewhat different lines adapted to our purposes).

\subsection{Lorentzian case}
We recall that the standard $4\times 4$ $\gamma$-matrices  fulfill,
\be
 \gamma_{\mu} \gamma_{\nu} +  \gamma_{\nu} \gamma_{\mu} = 2 \eta_{\mu\nu}
\ee
where $\eta_{\mu\nu}$ is the Minkowskian metric (chosen to have signature $(-+++)$.  We adopt a Majorana representation where the $\gamma$-matrices are real, with $\gamma_{k}$ symmetric and $\gamma_{0}$ antisymmetric.   Majorana spinors have then real components. Conjugate spinors $\bar \psi$ are defined by $\bar \psi = i \psi^\dagger \gamma^0$.  Also $\gamma^{\mu\nu} = \gamma^{[\mu}\gamma^{\nu]}$, etc

Under homogenous Lorentz transformations, Lorentz spinors transform as 
\be
\delta \psi =  \frac12  \omega_{\lambda \mu}\Sigma^{\lambda \mu} \psi
\ee
with
\begin{equation}
\Sigma^{\lambda \mu} = \frac14 [\gamma^\mu, \gamma^\mu]\, .
\end{equation}

The matrices $\Sigma^{\lambda \mu}$ provide a representation of the Lorentz algebra,
\begin{equation}
[\Sigma^{\lambda \mu}, \Sigma^{\rho \sigma}] = - \eta^{\lambda \rho} \Sigma^{\mu \sigma} + \eta^{\lambda \sigma} \Sigma^{\mu \rho} + \eta^{\mu \rho} \Sigma^{\lambda \sigma} - \eta^{\mu \sigma} \Sigma^{\lambda \rho} \, .
\end{equation}

\subsection{Trivial representations of the boosts}

We denote the flat Carroll metric and volume form respectively by  $\theta_{\mu \nu}$ and $\Omega$,
\begin{equation}
(\theta_{\mu \nu}) = \begin{pmatrix} 0 & 0 \\ 0 & I \end{pmatrix} \, , \quad (n^\mu) = \begin{pmatrix} 1 \\0 \\ 0 \\0  \end{pmatrix} \, , \quad \Omega = 1 \, . 
\end{equation}

The Carroll transformations on vectors are
\begin{equation}
\delta v^0 = b_k v^k \, , \qquad \delta v^k = {\omega^k}_m v^m \, ,
\end{equation}
($b_k$: boost parameters, ${\omega^k}_m$ with $\delta_{lk}{\omega^k}_m = - \delta_{mk}{\omega^k}_l $: spatial rotation parameters).

For covectors $\alpha_\mu$,
\begin{equation}
\delta \alpha_0 = 0 \, , \qquad \delta \alpha_k = -b_k \alpha_0 - \alpha_m {\omega^m}_k  \, ,
\end{equation}
so that $\alpha_\mu v^\mu$ is invariant.

These rules simplify to
\begin{equation}
\delta v^0 =0 \, , \qquad \delta v^k = 0 
\end{equation}
for longitudinal vectors ($v^\mu \theta_{\mu \nu} = 0 \Leftrightarrow v^\mu = \lambda n^\mu \Leftrightarrow v^k=0$) and
\begin{equation}
\delta \alpha_0 = 0 \, , \qquad \delta \alpha_k = - \alpha_m {\omega^m}_k 
\end{equation}
for transverse covectors ($\alpha_\mu n^\mu = 0 \Leftrightarrow \alpha_0 = 0$).  Both the spaces of longitunal vectors and transverse covectors are invariant.  

Transverse covectors do not transform under boosts.  One way to understand this property is to observe that the homogeneous Carroll algebra ${\mathcal C}$ is the semi-direct sum of the rotation algebra $so(3)$ and the abelian ideal $\mathcal B$ generated by the Carroll boosts. Thus, 
\begin{equation}
so(3) \simeq \frac{\mathcal C}{\mathcal B}
\end{equation}
and any representation of $so(3)$ can be trivially lifted to a representation of the Carroll group by declaring that boosts have a trivial action.   Transverse covectors define precisely a representation of this type.

One may write the transformation $\delta \alpha_k = - \alpha_m {\omega^m}_k$ in matrix notation for $\alpha \equiv (\alpha_k)$ as
\begin{equation}
\delta \alpha =\frac12 \omega_{lm}V^{lm} \alpha
\end{equation}
where the matrices $V^{lm}$ are equal to $- (R^{lm})^T$, with
\begin{equation}
{(R^{lm})^k}_r = \delta^{lk}\delta^{m}_r -\delta^{mk}\delta^{l}_r
\end{equation}
One has
\begin{equation}
[R^{lm}, R^{pq}] = - \delta^{lp} R^{mq} + \delta^{lq} R^{mp} + \delta^{mp} R^{lq} - \delta^{mq} R^{lp}
\end{equation}
expressing the $so(3)$ algebra (and similar relations hold for $[V^{lm}, V^{pq}]$).

One observes by duality that the quotient space of vectors modulo multiples of $n^\mu$ also define a representation on which the boosts act trivially,
\begin{equation}
\delta [v] =\frac12 \omega_{lm}R^{lm} [v]
\end{equation}
where $[v]$ is the equivalence class of the vectors $v^\mu$, completely characterized by its spatial components $v^k$.

Motivated by the transformation laws (\ref{Eq:TransSO(3)}) and  (\ref{Eq:InertBoost}) of the Carroll supercharge, we define a spinor $\psi$ as transforming only under $so(3)$, in the spin-$\frac12$ representation, i.e. \cite{Henneaux:1982qpq},

\begin{equation}
\delta \psi =  \frac12  \omega_{lm}\Sigma^{lm} \psi  \label{Eq:TransCSpin}
\end{equation}
where the matrices $\Sigma^{lm}$ are the generators of the spinor representation of $so(3)$ and 
fulfill the same commutation relations as the $R^{lm}$,
\begin{equation}
[\Sigma^{lm}, \Sigma^{pq}] = - \delta^{lp} \Sigma^{mq} + \delta^{lq} \Sigma^{mp} + \delta^{mp} \Sigma^{lq} - \delta^{mq} \Sigma^{lp}
\end{equation}

The spinors considered here, which do not transform under boosts, correspond to the electric spinors of \cite{Bergshoeff:2024ytq}.  They will be sufficient for our purpose of describing both the electric and magnetic limits of supergravity.

We now express their transformation rules under the Carroll group in terms of (Carrollian) $\gamma$-matrices.

\subsection{Carroll gamma matrices}

We introduce two sets of Carroll ``gamma" matrices $\{\Gamma^\mu\}$ and  $\{\rho_\mu\}$ through the relations:
\begin{eqnarray}
&& \Gamma^\mu \Gamma^\nu + \Gamma^\nu \Gamma^\mu =-2n^\mu n^\nu \, , \qquad \theta_{\mu \nu} \Gamma^\nu = 0 \, , \\
&& \rho_\mu \rho_\nu + \rho_\nu \rho_\mu =2 \theta_{\mu \nu} \, , \qquad \rho_\mu n^\mu = 0 \\
&& \Gamma^\mu \rho_\nu + \rho_\nu \Gamma^\mu = 0 \, .
\end{eqnarray}

From the first line we get,
\begin{equation}
\Gamma^\mu = \Gamma n^\mu
\end{equation}
for some matrix $\Gamma$ such that
\begin{equation}
\Gamma^2 = -I \, .
\end{equation}
Also,
\begin{equation}
\Gamma^\mu \rho_\mu = 0 \, .
\end{equation}

An explicit realization in terms of the Minkowskian gamma matrices is:
\begin{equation}
\Gamma^0 = \gamma^0 \equiv - \gamma_0\, , \quad \Gamma^k = 0 \, , \quad \rho_0 = 0 \, , \quad \rho_k = \gamma_k \, ,
\end{equation}

Because $\rho_\mu$ is transverse, there exist matrices $\rho^\mu$ such that $\theta_{\mu \nu} \rho^\nu = \rho_\mu$. The matrices $\rho^\mu$ are defined up to the addition of a matrix multiple of $n^\mu$, $\rho^\mu \rightarrow \rho^\mu + S n^\mu$. This ambiguity is irrelevant if $\rho^\mu$ is contracted with a transverse tensor.  In a Carroll frame, this ambiguity affects only $\rho^0$ but has no effect on $\rho^a$, equal to $\gamma^a$.

The transformation rule of a $4$-component Carroll spinor is as in (\ref{Eq:TransCSpin}), with 
\be
\Sigma^{ab} = \frac14 [\rho^a, \rho^b]\, .
\ee
(Thus, $\Sigma^{ab}_{Carroll} = \Sigma^{ab}_{Lorentz}$.)

The longitudinal $\Gamma^\mu$'s are numerically invariant
\begin{equation}
\delta \Gamma^0 = \frac12  \omega_{lm}[\Sigma^{lm}, \Gamma^0] = 0 \, , \quad \delta \Gamma^k =0 \, ,
\end{equation}
and similarly, the transverse $\rho_\mu$'s are also numerically invariant,
\begin{equation}
\delta \rho_k = \frac12  \omega_{lm}[\Sigma^{lm},\rho^k] - \rho_m {\omega^m}_k= 0 \, , \quad \delta \rho_0 =0 \, .
\end{equation}
[If we denote spinor indices by capital Latin letters, $\psi = (\psi^A)$, then the gamma-matrices $\Gamma^\mu$ read, e.g. ${{\Gamma^\mu}^A}_B$, and transform as a mixed vector-spinor-cospinor object. Similarly, the $\gamma_\mu$ transform as a mixed covector-spinor-cospinor object.]

\section{Tetrads and spinor fields on a pseudo-Riemmanian manifolds}
\label{Sec:LorentzianSpinors}

\subsection{Tetrads}

In order to describe spinor fields, we introduce in the tangent space of  each point of the manifold an orthornomal basis, which we denote by $h_{(\alpha)}$.  To avoid confusion with indices in the natural frame $\frac{\partial}{\partial x^\mu}$, frame indices are put in parentheses.

Since the vectors ${h_{(\alpha)}}^\mu$ are orthonormal one has
\be
g_{\lambda \mu} {h_{(\alpha)}}^\lambda {h_{(\beta)}}^\mu = \eta_{(\alpha) (\beta)}
\ee
We denote the dual basis of $1$-forms
 by ${h^{(\alpha)}}_\mu$, 
\be {h^{(\alpha)}}_\mu {h_{(\alpha)}}^\nu= \delta^\nu_\mu\, , \qquad {h^{(\alpha)}}_\mu {h_{(\beta)}}^\mu= \delta^{(\alpha)}_{(\beta)} \, ,
\ee
and also refer to the ${h^{(\alpha)}}_\mu$ as `` tetrads".
The metric can be expressed in terms of the tetrads
\be
 g_{\mu \nu} = \eta_{(\alpha) (\beta)} {h^{(\alpha)}}_\mu {h^{(\beta)}}_\nu \, ,
\ee
The tetrads are determined by the metric up to a local Lorentz transformation, which reads in infintesimal form,
\be
\delta {h^{(\alpha)}}_\lambda = \frac12 \omega_{(\alpha)(\beta)}{( L^{(\alpha)(\beta)})^{(\gamma)}}_{(\delta)} {h^{(\delta)}}_\lambda = {\omega^{(\alpha)}}_{(\beta)} {h^{(\beta)}}_\lambda
\ee
where
\be
{( L^{(\alpha)(\beta)})^{(\gamma)}}_{(\delta)} = \eta^{(\alpha)(\gamma)} \delta^{(\beta)}_{(\delta)} - \eta^{(\beta)(\gamma)} \delta^{(\alpha)}_{(\delta)}
\ee
are the (infinitesimal)  Lorentz matrices in the vector representation.

\subsubsection*{Time gauge}

It is convenient to parametrize the spacetime metric in terms of the lapse $N$, the shift $N^k$, and the spatial metric $g_{ij}$, and  to adopt the time gauge \cite{Dirac:1962} where the $(0)$-th leg of the tetrad $ {h_{(0)}}^\mu$  coincides with the normal to the constant $x^0$ hypersurfaces.  This yields
\begin{equation}
{h^{(0)}}_0 = N \, , \quad {h^{(0)}}_m = 0 \, , \quad {h^{(a)}}_0 = N^k{h^{(a)}}_k
\end{equation}
so that one can make the change of variables $({h^{(\lambda)}}_\mu) $ (in the time gauge) $\rightarrow (N, N^k, {h^{(a)}}_k)$
The spatial metric reads
\begin{equation}
g_{l m} = {h^{(a)}}_l \, {h^{(b)}}_m  \delta_{(a) (b)} \, .
\end{equation}
One has also
\begin{equation}
{h_{(0)}}^0 = \frac{1}{N} \, , \quad {h_{(0)}}^m = - \frac{N^m}{N} \, , \quad {h_{(a)}}^0 = 0 \, , \; ^{4 \hspace{-.05cm}} {h_{(a)}}^m = {h_{(a)}}^m \, .
\end{equation}
One can recover the behaviour of the fields under local Lorentz boosts at any given point by performing a surface deformation that induces such a transformatoin of the local frames  at the point ($N=0$ at the point but $N_{,k} \not= 0$).

\subsection{Covariant derivatives}

The $4$-dimensional spin connection
$
^{4 }\omega_{(\alpha)(\beta)\lambda} 
$
 without torsion is defined through
\begin{eqnarray}
&& \hspace{-1cm} ^{4 \hspace{- .08cm} }D_\lambda {h^{(\gamma)}}_{\rho}=  {h^{(\gamma)}}_{\rho  ;  \lambda} + \frac12 \, ^{4 }\omega_{(\alpha)(\beta)\lambda}  {(L^{(\alpha)(\beta)})^{(\gamma)}}_{(\delta)}{h^{(\delta)}}_{\rho} \equiv {h^{(\gamma)}}_{\rho  ;  \lambda} +  {^{4 }\omega^{(\gamma)}}_{(\beta)\lambda}  {h^{(\beta)}}_{\rho}  = 0 \,  , \\
&& {h^{(\gamma)}}_{\rho  ;  \lambda}  = \partial_\lambda {h^{(\gamma)}}_{\rho  }  - ^{4 \hspace{- .08cm} }{\Gamma^{\sigma}} _{\rho \lambda}  {h^{(\gamma)}}_{\sigma} \, ,
\end{eqnarray}
and is explicitly given by
\be
^{4 }\omega_{(\alpha)(\beta)\lambda}   = \frac12 \left({h_{(\alpha)}}^{\rho} h_{(\beta) \rho ; \lambda} - {h_{(\beta)}}^{\rho} h_{(\alpha) \rho ; \lambda}  \right)
\ee

The covariant derivatives of the gravitino field are
\be
^{4 \hspace{- .08cm} }D_\lambda \psi_\rho = \partial_\lambda \psi_{\rho  }  + \frac12 \, ^{4 }\omega_{(\alpha)(\beta)\lambda}  \Sigma^{(\alpha)(\beta)} \psi_{\rho}  - ^{4 \hspace{- .08cm} }{\Gamma^{\sigma}} _{\rho \lambda} \psi_{\sigma} \, ,
\ee
or if we replace the spacetime covector index $\rho$ by an internal index,
\be
^{4 \hspace{- .08cm} }D_\lambda \psi_{(\gamma)} = \partial_\lambda \psi_{(\gamma)  }  + \frac12 \, ^{4 }\omega_{(\alpha)(\beta)\lambda}  \Sigma^{(\alpha)(\beta)} \psi_{(\gamma)}  - \psi_{(\delta)} {^{4 }\omega^{(\delta)}}_{(\gamma)\lambda} \, ,
\ee

\subsection{Decomposition in space and time}

In the time gauge, one finds the following formulas connecting the spacetime connection to $3$-dimensional objects in a $3+1$ decomposition  \cite{Nelson:1978ex,Henneaux:1978wlm}:
\be
^{4 }\omega_{(a)(b)k} =  \omega_{(a)(b)k}
\ee
where $\omega_{(a)(b)k}$  is the $3$-dimensional spin connection without torsion defined through
\be
D_k {h^{(c)}}_{r}=  {h^{(c)}}_{r \vert k} + \frac12 \, \omega_{(a)(b)k} {(L^{(a)(b)})^{(c)}}_{(d)}{h^{(d)}}_{r} = 0 \,  , \quad {h^{(c)}}_{r \vert k} = \partial_k {h^{(c)}}_{r } - {\Gamma^{s}} _{r k}  {h^{(c)}}_{s} \, ,
\ee
given by
\be
\omega_{(a)(b)k}   = \frac12 \left({h_{(a)}}^{r} h_{(b) r  \vert k} - {h_{(b)}}^{r}{h_{(a)}}_{r \vert k} \right) \, .
\ee
One also finds
\be
^{4 }\omega_{(0)(b)k} =  K_{km} {h_{(b)}}^{m}
\ee

It follows that 
\be
^{4\hspace{-.05cm} }D_k \psi_{(c)} = D_k \psi_{(c)} + K_{kn} {h_{(b)}}^{n} \, \Sigma^{(0)(b)} \psi_{(c)} + \psi_{(0)} K_{kn} {h_{(c)}}^{n}
\ee
Similarly,
\be
^{4\hspace{-.05cm} }D_k \psi_{(0)} = D_k \psi_{(0)} + K_{kn} {h_{(b)}}^{n} \, \Sigma^{(0)(b)} \psi_{(0)} + \psi^{(c)} K_{kn} {h_{(c)}}^{n}
\ee

A straightforward computation gives on the other hand\footnote{$ \Delta_0$ is denoted $\Delta_N$ in \cite{Henneaux:1978wlm}.}
\be
N \, ^{4\hspace{-.05cm} }D_{(0)} \psi_{(c)} = \Delta_0 \psi_{(c)} - \partial_n N  \Sigma^{(0)(b)} \psi_{(c)}{h_{(b)}}^{n} - \psi_{(0)}  \partial_n N  {h_{(c)}}^{n}
\ee
with
\be
\Delta_0 \psi_{(c)}  = \partial_0 \psi_{(c)} + \frac12 \Xi_{(a)(b)} \Sigma^{(a)(b)} \psi_{(c)} - \psi_{(a)} {\Xi^{(a)}}_{(c)} -\mathcal L_{N^k} \psi_{(c)}  \label{Eq:DefDelta}
\ee
and
\be
 \Xi_{(a)(b)} = \frac12 \left({h_{(a)}}^{r} \partial_0 h_{(b) r } - {h_{(b)}}^{r} \partial_0 h_{(a)r} \right)  \label{Eq:DefXi}
\ee
Here, $\mathcal L_{N^k} \psi_{(c)}$ is the three-dimensional Lie derivative of $ \psi_{(m)}$ along the shift $3$-vector $N^k$,
\be
\mathcal L_{N^k} \psi_{(c)}  = N^k\partial_k \psi_{(c)} + \frac12 \Theta_{(a)(b)} \Sigma^{(a)(b)} \psi_{(c)} - \psi_{a} {\Theta^{(a)}}_{(c)} 
\ee
with
\be
\Theta_{(a)(b)} =\frac12 \left({h_{(a)}}^{r} \, l_{N^k} h_{(b) r } - {h_{(b)}}^{r} \,  l_{N^k} h_{(a)r} \right) \label{Eq:DefTheta}
\ee
(see Appendix {\bf \ref{Sec:LieDerivatives}}).

\subsection{Supersymmetry in the time gauge}
\label{Subsec:SusyTimeGauge}

The supersymmetry transformation can also be decomposed into space and time as follows in the time gauge,
\begin{eqnarray}
&&\delta {h^{(a)}}_m = \frac12 \kappa^2 \bar \epsilon \gamma^{(a)} \psi_m \, , \\
&&\delta N =  \frac{N}{2} \kappa^2 \bar \epsilon \gamma^{(0)} \psi_{(0)} \, , \\
&&\delta N^k =\frac{N}{2} \kappa^2 \left(\bar \epsilon \gamma^{(a)} \psi_{(0)} + \bar \epsilon \gamma_{(0)} \psi^{(a)}\right){h_{(a)}}^k \, , \\
&& \delta \psi_{(a)} =  D_{(a)} \epsilon + K_{kn} {h_{(a)}}^k {h_{(b)}}^n \Sigma^{(0)(b)} \epsilon + \kappa^2 \epsilon \psi^2\, , \\
&& \delta \psi_{(0)} = \frac{1}{N}\Delta_0 \epsilon -   \frac{\partial_n N}{N} \Sigma^{(0)(a)} \epsilon {h_{(a)}}^n+ \kappa^2 \epsilon \psi^2 \, , 
\end{eqnarray}
where we have included the Lorentz boost needed to maintain the time gauge - which redefines also the $\kappa^2 \epsilon \psi^2$-terms.

\section{Carrollian tetrads}
\label{Sec:CarrollTetrads}

\subsection{Properties}

At each point of a Carrollian manifold, one can introduce a Carrollian frame, i.e., an oriented basis of tangent vectors $\{ {h_{(\alpha)}}^\mu\}$ in which the Carrollian metric and null vector at that point takes the canonical form, that is
\begin{equation}
{h_{(\alpha)}}^\mu {h_{(\beta)}}^\nu g_{\mu \nu}  = \theta_{(\alpha) (\beta)} \, , \quad  (\theta_{(\alpha) (\beta)})= \begin{pmatrix} 0 & 0 \\ 0 & \delta_{(a)(b)} \end{pmatrix} \, , \qquad  \det \left({h_{(\alpha)}}^\mu \right) \Omega = 1
\end{equation}
(again, indices in parentheses label the vectors, indices not in parentheses are spacetime components).  This implies
\begin{equation}
{h_{(0)}}^\mu = n^\mu
\end{equation}
Carrollian frames, or ``Carrollian tetrads", are defined up to a homogeneous Carrollian transformation, 
\begin{eqnarray}
&& {h_{(0)}}^\mu \rightarrow {h'_{(0)}}^\mu  = {h_{(0)}}^\mu \\
&& {h_{(a)}}^\mu \rightarrow {h'_{(a)}}^\mu  = {h_{(b)}}^\mu   {(R^{-1})^{(b)}}_{(a)}  - b_{(a)}{h_{(0)}}^\mu\end{eqnarray}
where $ {R^{(a)}}_{(b)}$ defines a spatial rotation and $b_{(a)}$ a Carroll boost. (If $ {R^{(a)}}_{(b)}$ acts on internal vectors, ${(R^{-1})^{(b)}}_{(a)}$ acts on internal covectors.)

Since the $\{ {h_{(\alpha)}}^\mu\}$ form a basis, the corresponding matrix can be inverted. The dual basis is denoted by $\{ {h^{(\alpha)}}_\mu\}$ and fulfills
\begin{equation}
{h^{(\alpha)}}_\mu \, {h_{(\beta)}}^\mu ={\delta^{(\alpha)}}_{(\beta)}
\end{equation}
as well as
\begin{equation}
{h^{(\alpha)}}_\mu \, {h_{(\alpha)}}^\lambda ={\delta^{\lambda}}_{\mu}
\end{equation}
(right and left inverses coincide).

It follows from the above relations that 
\begin{equation}
g_{\lambda \mu} = {h^{(\alpha)}}_\lambda \, {h^{(\beta)}}_\mu  \theta_{(\alpha) (\beta)} \, , \qquad  \Omega =  \det \left({h^{(\alpha)}}_\mu \right) \label{Eq:MetricOmega}
\end{equation}
Carroll transformations act on the dual basis as follows
\begin{eqnarray}
&& {h^{(0)}}_\mu \rightarrow {h'^{(0)}}_\mu  = {h^{(0)}}_\mu + b_{(a)}{h^{(a)}}_\mu  \label{Eq:CarrollT1}\\
&& {h^{(a)}}_\mu \rightarrow {h'^{(a)}}_\mu  = {R^{(a)}}_{(b)} {h^{(b)}}_\mu \label{Eq:CarrollT2}
\end{eqnarray}
When no risk of confusion can arise, we will also call the dual basis ``(Carrollian) tetrads".

Note that a choice of local Carroll frame implies a choice of Ehresmann connection \cite{Ciambelli:2019lap}.

\subsection{Time gauge}

Given a slicing of spacetime by hypersurfaces of constant $x^0$ transverse to the null vector field $n^\mu$, one can choose the tetrads so that the three spatial legs are tangent to the hyersurfaces.  Such a condition reduces the local gauge freedom from Carroll transformations to spatial rotations.  The corresponding gauge is called the ``time gauge" and define ``adapted tetrads" to the foliation.  It reads
\begin{equation}
{h_{(a)}}^0 = 0
\end{equation}
or equivalently, in terms of the dual basis,
\begin{equation}
{h^{(0)}}_m = 0
\end{equation}

We now parametrize the metric in terms of the Carroll lapse and the shift, defined by
\begin{equation}
\frac{\partial}{\partial x^0} = N \, n + N^k \frac{\partial}{\partial x^k} \quad \Leftrightarrow \quad n^0 = \frac{1}{N} \, , \quad n^k = - \frac{N^k}{N}
\end{equation}
i.e.,
\begin{equation}
(g_{\alpha \beta}) = \begin{pmatrix}N^k N_k & N_b \\ N_a & g_{ab} \end{pmatrix} \, , \quad \Omega = N \sqrt{g}  \label{Eq:Parametrization}
\end{equation}
where indices are lowered and raised with the spatial metric.  We stress that $N$ is here the Carroll lapse, i.e., either $N^E$ or $N^M$ depending on the context.

One then finds
\begin{equation}
{h_{(0)}}^0 = \frac{1}{N} \, , \quad {h_{(0)}}^m = - \frac{N^m}{N} \, , \quad {h_{(a)}}^0 = 0
\end{equation}
and
\begin{equation}
{h^{(0)}}_0 = N \, , \quad {h^{(0)}}_m = 0 \, , \quad {h^{(a)}}_0 = N^k{h^{(a)}}_k
\end{equation}
and the $3\times 3$ matrix $({h_{(a)}}^k)$ is inverse to the $3\times 3$ matrix $({h^{(b)}}_m)$ (i.e., $\, {\!^3{h_{(a)}}}^k = \!^4{h_{(a)}}^k$).
The spatial metric simply reads
\begin{equation}
g_{l m} = {h^{(a)}}_l \, {h^{(b)}}_m  \delta_{(a) (b)} \, .
\end{equation}

In the time gauge, one can equivalently replace the independent components of $ {h^{(\alpha)}}_\mu$ by the variables $ {h^{(a)}}_m$, $N$ and $N^k$ in the variational principle.

It is instructive to compare the parametrization (\ref{Eq:Parametrization}) of the metric with the corresponding parametrization in the Lorentzian case, which is
\begin{equation}
(g_{\alpha \beta}) = \begin{pmatrix}- (N_{Lorentz})^2 + N^k N_k & N_b \\ N_a & g_{ab} \end{pmatrix} \, .  \label{Eq:ParametrizationR}
\end{equation}
In both the electric and magnetic cases, the Lorentzian lapse goes to zero as we take the Carrollian limit.

\section{Lie derivatives of Carroll spinor fields}
\label{Sec:LieDerivatives}

Lie derivatives of spinor fields on a Riemannian manifolds have been defined in \cite{Kosmann:1971ugf,SSH:1977}.  We extend here the definition to Carrollian manifolds.

Lie derivatives of tensor fields are natural in the sense that they do not involve any extra structure besides the differential structure of the manifold and hence do not depend on whether the manifold is Riemannian or Carrollian.  This is not so for spinor fields - more generally, for objects transforming under the local Lorentz or Carroll group of the tangent space -  if one insists, as one should, that their Lie derivatives possess the same local transformation properties under local transformations as the fields themselves, without derivatives of the local gauge parameter. What makes the question not entirely trivial with respect to, say, covariant derivatives, is that in general, the ordinary Lie derivatives of the tetrads (viewed as tangent vectors) along an arbitrary vector field define a general linear transformation, and not a local Lorentz or Carroll transformation, unless the vector field in question corresponds to an isometry of the structure. 

This raises the question of extracting an infinitesimal Lorentz or Carroll transformation from a general linear transformation of the tetrads.

The procedure adopted in the Riemannian case goes as follows \cite{Kosmann:1971ugf,SSH:1977}.  We denote $l_\xi  {h^{(\alpha)}}_\mu$ the ordinary Lie derivatives of the tetrads viewed as individual one-forms and define the complete Lie derivative $\mathcal L _\xi {h^{(\alpha)}}_\mu$ by adding a local Lorentz rotation,
\be 
\mathcal L _\xi {h^{(\alpha)}}_\mu = l_\xi  {h^{(\alpha)}}_\mu + {\Omega^{(\alpha)}}_{(\beta) \, \xi} {h^{(\beta)}}_\mu
\ee
such that 
\be
\mathcal L _\xi {h^{(\alpha)}}_\mu =  \frac12  h^{(\alpha)\nu}\mathcal L _\xi g_{\mu \nu}
\ee
This choice guaranteees that $\mathcal L _\xi {h^{(\alpha)}}_\mu = 0$ when $\mathcal L _\xi g_{\mu \nu} = 0$.  The derivative $l_\xi  {h^{(\alpha)}}_\mu$ is the Lie derivative analog of the incomplete covariant derivative $  {h^{(\alpha)}}_{\mu; \lambda}$ while $\mathcal L _\xi {h^{(\alpha)}}_\mu$ corresponds to $D_\lambda {h^{(\alpha)}}_\mu$.
By direct computation, one finds
\be
\Omega_{(\alpha)(\beta) \xi} = \frac12 \left( {h_{(\alpha)}}^\mu \, l_\xi  h_{(\beta) \mu} - {h_{(\beta)}}^\mu \, l_\xi  h_{(\alpha) \mu}\right)  \label{Eq:LieDerivativeConnection}
\ee

The Lie derivative of a spinor field is then defined by adding the same Lorentz transformation term to $l_\xi \psi =  \xi^\mu \partial_\mu \psi$, i.e., 
\be
\mathcal L_\xi \psi = l_\xi \psi + \frac12 \Omega_{(\alpha)(\beta) \xi}  \Sigma^{(\alpha)( \beta)} \psi \, .
\ee
The Lie derivative $\mathcal L_\xi \psi$ transforms as a spinor under local Lorentz rotations, while $\xi^\mu \partial_\mu \psi$ does not.   The Lie derivative $\mathcal L_\xi \psi$ is thus the natural quantity to use to build covariant expressions out of the Lie derivatives of fields.

Similarly, one defines the Lie derivative of the gravitino field,
\be
\mathcal L_\xi \psi_{(\gamma)} = l_\xi \psi + \frac12 \Omega_{(\alpha)(\beta) \xi}  \Sigma^{(\alpha)( \beta)} \psi_{(\gamma)} + \psi_{(\alpha)}  {\Omega^{(\alpha)}}_{(\gamma) \, \xi}\, .
\ee

One proceeds in the Carroll case along exactly the same lines.   Since our local fields are  inert under Carroll boosts, one only needs to extract the rotation associated with the Lie derivatives of the Carroll tetrads, which actually just takes the same expression as (\ref{Eq:LieDerivativeConnection}) with indices $\alpha$, $\beta$ specialized to spatial values
\be
\Omega_{(a)(b) \xi} = \frac12 \left( {h_{(a)}}^\mu \, l_\xi  h_{(b) \mu} - {h_{(b)}}^\mu \, l_\xi  h_{(a) \mu}\right)
\ee
We thus define
\be
\mathcal L_\xi \psi = l_\xi \psi + \frac12 \Omega_{(a)(b) \xi}  \Sigma^{(a)( b)} \psi \, .
\ee
In particular,
\begin{eqnarray}
&&\mathcal L_n \psi =n^\mu \partial_\mu \psi + \frac12 \Omega_{(a)(b) }  \Sigma^{(a)( b)} \psi \,  , \\
&& \Omega_{(a)(b) } = \frac12 \left( {h_{(a)}}^\mu \, l_n  h_{(b) \mu} - {h_{(b)}}^\mu \, l_n  h_{(a) \mu}\right) = \frac{1}{N}\left(\Xi_{(a)(b)} - \Theta_{(a)(b)} \right)
\end{eqnarray}
(see (\ref{Eq:DefXi}) and (\ref{Eq:DefTheta})).
For a transverse vector-spinor $\Phi_{(\alpha)}$ ($\Phi_{(0)}=0$),
\be
\mathcal L_n \Phi_{(c)} =n^\mu \partial_\mu \Phi_{(c)}+  \frac12 \Omega_{(a)(b) }  \Sigma^{(a)( b)}  \Phi_{(c)} + \Phi_{(a)} {\Omega^{(a)}}_{(c)}\, , \qquad \mathcal L_n \Phi_{(0)}=0
\ee
(so that $\mathcal L_n \Phi_{(\alpha)}$ is not only a vector-spinor but is also transverse).

We close this appendix by noting that in the $3+1$ decomposition in the time gauge, where both Lorentz and Caroll groups reduce to the rotation subgroup, Lie derivatives along tangent vectors - such as the shift - of spinors coincide in the Lorentz and in the Carroll cases.

\end{document}